\documentclass[aps,prl,showpacs,twocolumn,groupedaddress,nofootinbib]{revtex4}
\usepackage{graphicx}
\usepackage{dcolumn} 
\usepackage{bm}      
\usepackage{amssymb} 
\usepackage{lineno}

\newcommand{\bsdecay}{B_{s}^{0}\to J/\psi\phi}
\newcommand{\bddecay}{B_{d}^{0}\to J/\psi K^{*0}}
\newcommand{\bd}{B_{d}^{0}}
\newcommand{\bs}{B_{s}^{0}}
\newcommand{\jpsi}{J/\psi }
\newcommand{\kstar}{K^{*0}}
\newcommand{\acero}{|A_{0}|^{2}}
\newcommand{\all}{|A_{\parallel}|^{2}}
\newcommand{\duno}{\delta_{1}}
\newcommand{\ddos}{\delta_{2}}
\newcommand{\dll}{\delta_{\parallel}}
\newcommand{\dt}{\delta_{\perp}}

\newcommand{\ds}{\delta_{s}}

\newcommand{\dg}{\Delta\Gamma_{s}}
\newcommand{\taus}{\bar{\tau}_{s}}

\newcommand{\tstd}{\bar{\tau}_{s}/\tau_{d}}

\begin{document}
\hspace{5.2in}\mbox{FERMILAB-PUB-08/418-E}
\title{Measurement of the angular and lifetime parameters of the decays \\
    {\boldmath $\bddecay$ \unboldmath} and {\boldmath $\bsdecay$ \unboldmath}}
%
\author{V.M.~Abazov$^{36}$}
\author{B.~Abbott$^{75}$}
\author{M.~Abolins$^{65}$}
\author{B.S.~Acharya$^{29}$}
\author{M.~Adams$^{51}$}
\author{T.~Adams$^{49}$}
\author{E.~Aguilo$^{6}$}
\author{M.~Ahsan$^{59}$}
\author{G.D.~Alexeev$^{36}$}
\author{G.~Alkhazov$^{40}$}
% ...

\affiliation{\vspace{0.1 in}(The D\O\ Collaboration)\vspace{0.1 in}}
% ...
\affiliation{$^{6}$University of Alberta, Edmonton, Alberta, Canada,
                Simon Fraser University, Burnaby, British Columbia, Canada,
                York University, Toronto, Ontario, Canada, and
                McGill University, Montreal, Quebec, Canada}
% ...
\affiliation{$^{29}$Tata Institute of Fundamental Research, Mumbai, India}
% ...
\affiliation{$^{36}$Joint Institute for Nuclear Research, Dubna, Russia}
% ...
\affiliation{$^{40}$Petersburg Nuclear Physics Institute,
                St. Petersburg, Russia}
% ...
\affiliation{$^{49}$Florida State University, Tallahassee, Florida 32306, USA}
% ...
\affiliation{$^{51}$University of Illinois at Chicago,
                Chicago, Illinois 60607, USA}
% ...
\affiliation{$^{59}$Kansas State University, Manhattan, Kansas 66506, USA}
% ...
\affiliation{$^{65}$Michigan State University,
                East Lansing, Michigan 48824, USA}
% ...
\affiliation{$^{75}$University of Oklahoma, Norman, Oklahoma 73019, USA}
% ...

\date{September 30, 2008}

\begin{abstract}

We present measurements of the linear polarization amplitudes and the strong 
relative phases that describe the flavor-untagged decays $\bddecay$ and $\bsdecay$ 
in the transversity basis. We also measure the mean lifetime $\taus$ of the $\bs$ 
mass eigenstates and the lifetime ratio $\tstd$. The analyses are based on approximately 
2.8 fb$^{-1}$ of data recorded with the D0 detector. From our measurements of the 
angular parameters we conclude that there is no evidence for a deviation from 
flavor SU(3) symmetry for these decays and that the factorization assumption is 
not valid for the $\bddecay$ decay.

\end{abstract}

\maketitle

\vskip.5cm

\newpage

$B$ mesons are fertile ground to study CP violation and search for evidence of
new physics. There are elements, in addition to CP violation, involved in
the theoretical description of $B$ meson decays, such as flavor SU(3) symmetry,
factorization and final-state strong interactions.  To understand the role
CP violation plays in these decays, it is essential to understand and
isolate the effect of each of these elements in the $B$ meson decays.

Factorization states that the decay amplitude of $B$ meson decays can be
expressed as the product of two single current matrix elements~\cite{browder} and 
this implies that the relative strong phases are 0 (mod $\pi$)~\cite{Fleischer}.
A different measured value for the strong phases would indicate the presence of 
final-state strong interactions. The $\bd$ meson can be formed by replacing the 
$s$ quark with the $d$ quark in the $\bs$ meson. From flavor SU(3) symmetry 
applied to the $\bd$-$\bs$ system one expects that the theoretical description 
is similar; in particular the $\bddecay$ and $\bsdecay$~\cite{charge.conjugation} 
decays, can be described in the transversity basis~\cite{Fleischer} by the linear 
polarization amplitudes, $A_{0},A_{\parallel},$ and $A_{\perp}$, and the relative 
strong phases $\duno$ and $\ddos$. Flavor SU(3) symmetry requires that the 
amplitudes and phases characterizing these decays should have the same values.

Other observables of these decays are the lifetimes of both mesons, which allow 
us to compare with theoretical predictions of the lifetime ratio. 
Phenomenological models predict differences of about $1\%$ 
\cite{theorylifetime,franco} between the $\bd$ and $\bs$ lifetimes. Previous $B$ 
meson lifetime measurements \cite{pdg1} are consistent with these predictions.

In this Letter we report the measurements of the parameters that describe
the time-dependent angular distributions of the decays $\bddecay$ and $\bsdecay$
in the transversity basis, where the initial $B$ meson flavor is not determined
(``untagged"). We study the $\bd$ and $\bs$ mesons to verify the validity of the 
factorization assumption~\cite{Fleischer} and to check if flavor SU(3) 
symmetry~\cite{Fleischer} holds for these decays.  We also report the lifetime 
ratio $\tstd$ for these mesons and the width difference $\dg$ between the light 
and heavy $\bs$ mass eigenstates. The analyses were performed using data collected 
with the D0 detector~\cite{run2det} in Run II of the Fermilab Tevatron Collider 
during $2003-2007$ with an integrated luminosity of approximately 
$2.8\mathrm{~fb}^{-1}$ of $p\bar{p}$ collisions at a center-of-mass energy of 
1.96 TeV.  In contrast with the flavor-tagged analysis reported in 
Ref.~\cite{tagged.d0}, in this Letter we report a simultaneous analysis of both 
the $\bd$ and $\bs$ meson decays, carried out in such a way that a 
straightforward comparison between their angular and lifetime parameters can be 
performed.

We use the $\bsdecay$, $\jpsi\rightarrow\mu^{+}\mu^{-}$, $\phi\rightarrow K^{+}K^{-}$
selection described in Ref. \cite{prl.daria}. The decay $\bddecay$, 
$\jpsi\rightarrow\mu^{+}\mu^{-}$, $K^{*0}~\rightarrow~K^{\pm}\pi^{\mp}$ is 
reconstructed using similar selection criteria and algorithms as the $\bs$ 
channel because they have the same four-track topology in the final state. The 
differences are the requirement that the transverse momentum of the pion be greater 
than $0.7\mathrm{~GeV}/c$, the invariant mass for the $(\jpsi,\kstar(892))$ pair be 
in the range $4.93-5.61\mathrm{~GeV}/c^{2}$, and the selection of the $\kstar(892)$ 
candidates by demanding the two-particle invariant mass between $850\mathrm{~MeV}/c^{2}$ 
and $930\mathrm{~MeV}/c^{2}$. Due to lack of charged particle identification, 
we assign the mass of the pion and kaon to the latter two tracks and use the 
combination with invariant mass closest to the $K^{*0}$ mass.

\begin{figure}[htbp]
\includegraphics[scale=0.35]{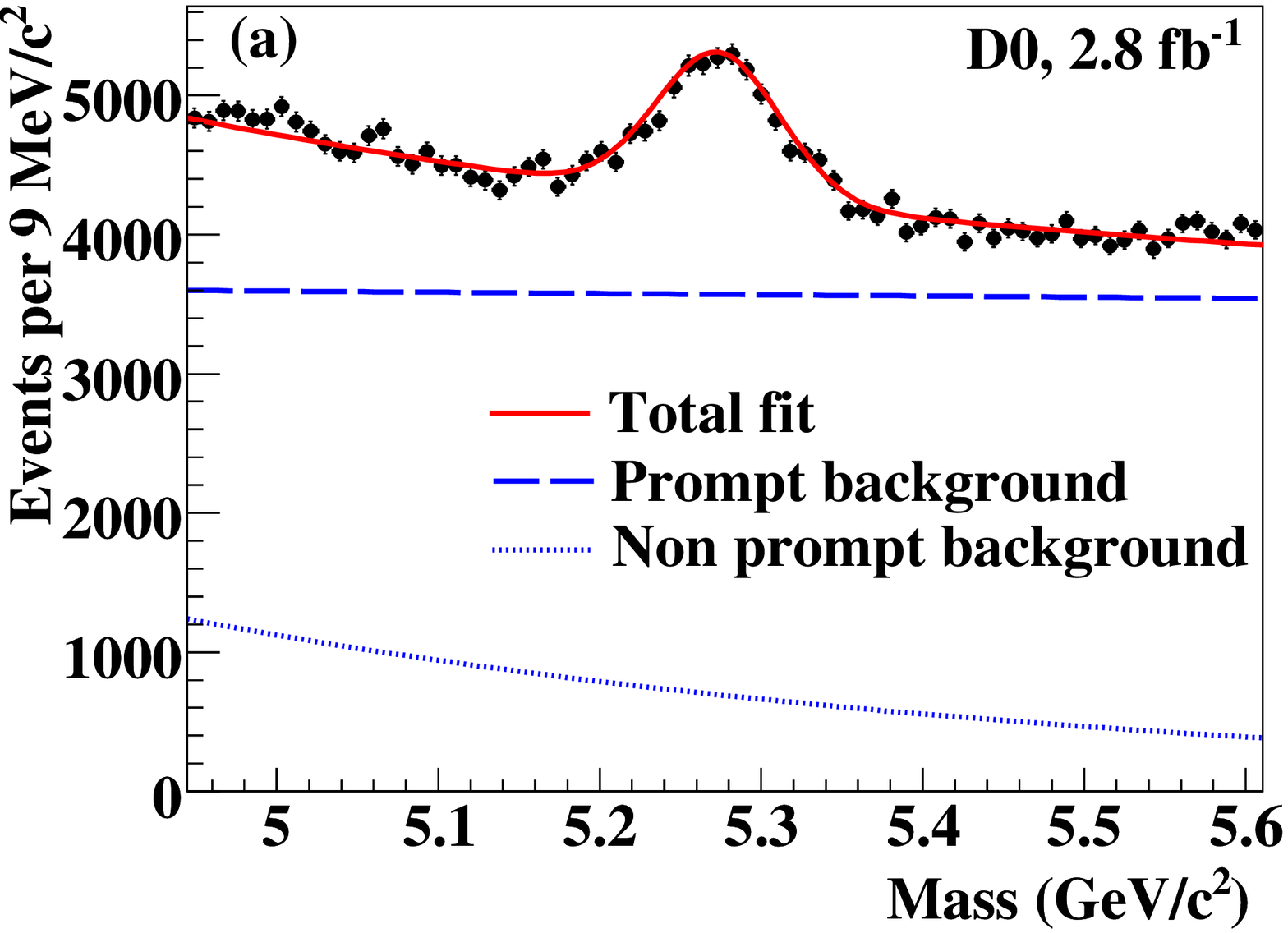}\\
\includegraphics[scale=0.35]{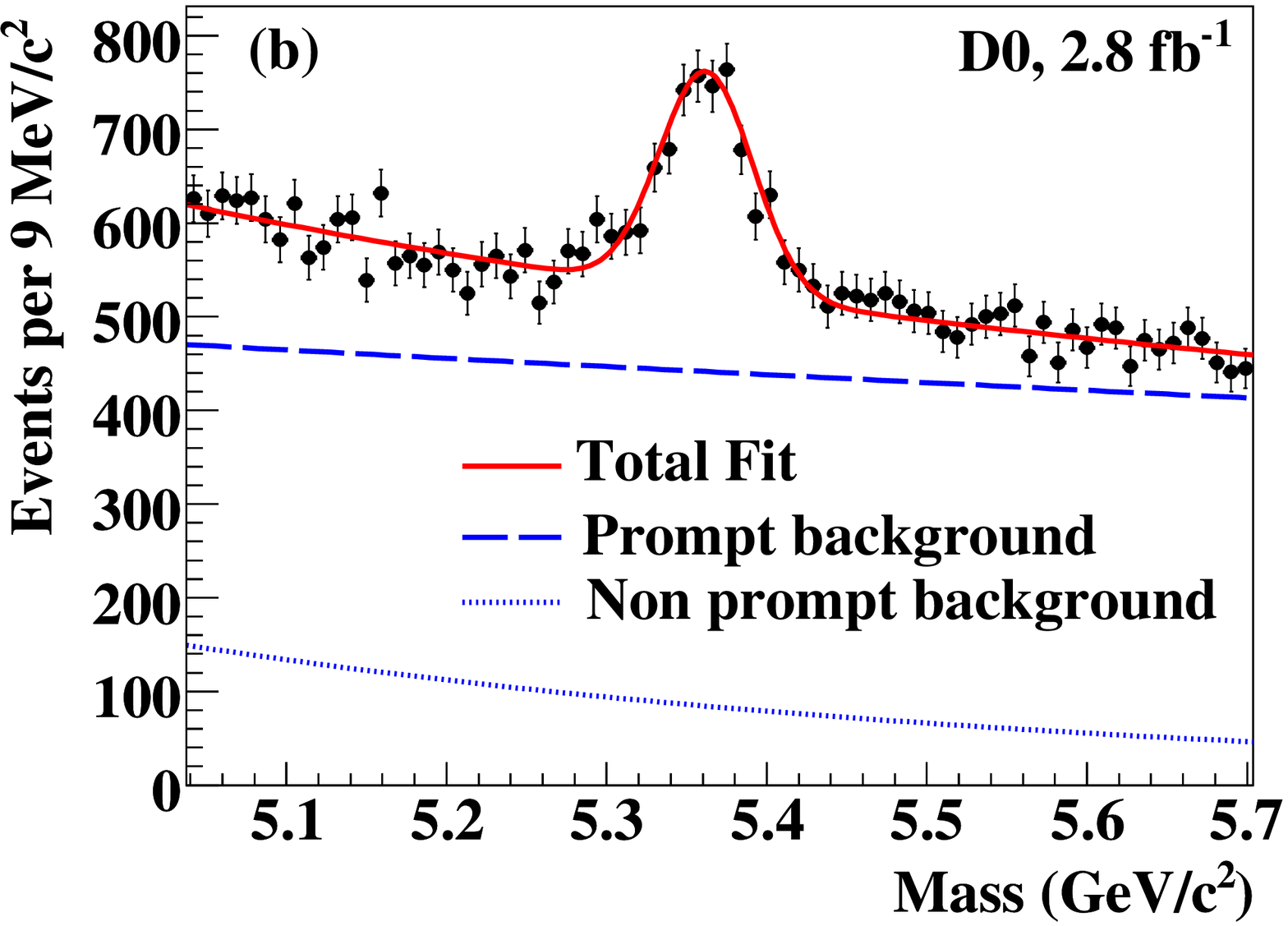}\\
\caption{ Invariant mass distribution for selected (a) $\bd$ and (b) $\bs$ 
candidate events. The points with error bars represent the data, and the curves 
represent the fit projections for the total and the background components.}\label{fig:mass.sample.a}
\end{figure}

The proper decay length (PDL), defined as in Refs.~\cite{prl.eduard,prl.pedro}, 
for a given $\bd$ or $\bs$ candidate is determined by measuring the distance 
traveled by each $b$-hadron candidate in a plane transverse to the beam direction, 
and then applying a Lorentz boost correction. In the $\bd$ and $\bs$ final 
selection, we require a PDL uncertainty of less than $60~\mu$m.  We find $334199$ 
and $41691$ candidates that pass the $\bd$ and $\bs$ selection criteria, respectively 
(see Fig.~\ref{fig:mass.sample.a}).

We denote the set of the angular variables defined in the transversity basis, 
where the decays $\bddecay$ and $\bsdecay$ are studied, as 
$\bm{\omega}=\left\{\varphi,\cos\theta,\cos\psi\right\}$. The description of
these decays in this basis gives us access to the three
linear polarization amplitudes at production time, $t=0$, $|A_{0}(0)|$, 
$|A_{\parallel}(0)|$, and $|A_{\perp}(0)|$, satisfying
$\acero+\all+|A_{\perp}|^{2}=1$~\cite{amplitudes}; and the CP-conserving strong 
phases $\duno\equiv\arg[A_{\parallel}^{*}A_{\perp}]$, and
$\ddos\equiv\arg[A_{0}^{*}A_{\perp}]$.
Since only the relative phases of the amplitudes can enter physics observables,
we are free to fix the phase of one of them, and we choose to fix
$\delta_{0}\equiv\arg(A_{0})=0$.

According to the standard model, CP-violation effects in the $\bs$ system are very 
small~\cite{lenz}. In this analysis, we assume CP conservation and express the
differential decay rate for the untagged decay $\bsdecay$ as~\cite{Fleischer}:
\begin{eqnarray}\label{decay.distrib.bs}
\nonumber d^{4}{\cal P}/\left(d\bm{\omega}\;dt\right) & \propto & e^{-\Gamma_{L}t}\left[\acero f_{1}(\bm{\omega}) + \mathrm{Re}(A_{0}^{*}A_{\parallel})f_{5}(\bm{\omega})\right.\\
&+&\left.\all f_{2}(\bm{\omega})\right] + e^{-\Gamma_{H}t}|A_{\perp}|^{2}f_{3}(\bm{\omega}),
\end{eqnarray}
where $\Gamma_{L(H)}\equiv1/\tau_{L(H)}$ is the inverse of the lifetime 
corresponding to the light (heavy) mass eigenstate. The measured parameters, the
width difference $\dg\equiv\Gamma_{L}-\Gamma_{H}$ and the mean lifetime 
$\taus\equiv1/\bar{\Gamma}=2/\left(\Gamma_{L}+\Gamma_{H}\right)$,
are given in terms of these inverse lifetimes. The angular functions 
$f_{i}(\bm{\omega})$ are defined in Ref.~\cite{Fleischer}.
In this decay, we have access to the phase $\dll=\arg(A_{0}^{*}A_{\parallel})$, 
which is related to $\duno$ and $\ddos$ by $\dll=\ddos-\duno$.

In the $\bd$ system, there is evidence of interference between the $P$- and 
$S$-wave $K\pi$ amplitudes~\cite{babar}, which is taken into account in this 
analysis. The differential decay rate for the untagged decay $\bddecay$ is given
by~\cite{Fleischer,babar}:
\begin{eqnarray}\label{decay.distrib.bd}
\nonumber d^{4}{\cal P}/\left(d\bm{\omega}\;dt\right) & \propto & e^{-\Gamma_{d}t}\left\{\cos^{2}\lambda\left[\acero f_{1}(\bm{\omega})+\all f_{2}(\bm{\omega})\right.\right.\\
\nonumber&+&\left.|A_{\perp}|^{2}f_{3}(\bm{\omega})-\zeta\;\mathrm{Im}(A_{\parallel}^{*}A_{\perp})f_{4}(\bm{\omega})\right.\\
\nonumber&+&\left.\mathrm{Re}(A_{0}^{*}A_{\parallel})f_{5}(\bm{\omega})+\zeta\;\mathrm{Im}(A_{0}^{*}A_{\perp})f_{6}(\bm{\omega})\right]\\
\nonumber&+&\sin^{2}\lambda\cdot f_{7}(\bm{\omega})\nonumber\\
\nonumber&+&\frac{1}{2}\sin2\lambda\left[f_{8}(\bm{\omega})\cos\left(\dll - \ds\right)|A_{\parallel}|\right.\nonumber\\
\nonumber&+&\left. f_{9}(\bm{\omega})\sin\left(\dt - \ds\right)|A_{\perp}|\right.\nonumber\\
&+&\left.\left.f_{10}(\bm{\omega})\cos\ds\cdot|A_{0}|\right]\right\},
\end{eqnarray}
where $\Gamma_{d}\equiv1/\tau_{d}$ is the inverse of the $\bd$ lifetime, 
$\zeta=+1(\zeta=-1)$ for $K^{+}(K^{-})$; $\lambda$, $\ds$, and $f_{i}(\bm{\omega})$ 
are defined in Refs.~\cite{Fleischer,babar}.  For the $\bd$, $\Delta\Gamma_{d}$ 
is expected to be zero~\cite{lenz}.

An unbinned likelihood fit is performed to extract all the $\bd$ and $\bs$ 
parameters. For the $j$th $B$ meson candidate, the inputs for the fit are the
mass $m_{j}$, PDL $ct_{j}$, PDL uncertainty $\sigma_{ct_{j}}$, and the angular
variables $\bm{\omega}_{j}$. The likelihood function ${\cal L}$ for the untagged 
decays $\bddecay$ and $\bsdecay$, is defined by
\begin{equation}
\begin{array}{lcl}
{\cal L}& = & \prod_{j=1}^N \left[f_{s}{\cal F}_{s}^{j}+(1-f_{s}){\cal F}_{b}^{j}\right],
\end{array}
\end{equation}
where $N$ is the total number of selected events and $f_s$ is the fraction of
signal events in the sample, a free parameter in the fit.

${\cal F}_{s}$ is the product of the signal probability distribution functions 
(PDF) of mass, PDL, and transversity angles, and the angular acceptances, which 
are determined via Monte Carlo simulations. The mass and PDL signal distributions 
are modeled for both decays in the same way. The mass distribution is modeled by 
a Gaussian function with free mean and width. The PDL distribution is 
described~\cite{prl.eduard} by the convolution of an exponential, whose decay 
constant is one of the fit parameters with a resolution function represented by 
two weighted Gaussian functions centered at zero. The widths $s_{i}\sigma_{ct_{j}}$ 
of each Gaussian with scale factors $s_{i}\thinspace\thinspace(i=1,2)$ are free
parameters in the fit to allow for a possible misestimate of the PDL uncertainty. 
The transversity angular distributions are modeled by the corresponding 
normalized Eqs. (\ref{decay.distrib.bs}) and (\ref{decay.distrib.bd}).  The 
contribution where the mass of the $K$ and $\pi$ are misassigned in our data is 
estimated by using Monte Carlo studies to be about $13\%$ and is taken into account.

${\cal F}_{b}$ is the product of the background PDF of the same
variables and the angular acceptance as in the signal.  We separate the
background contributions into two types. The prompt background
accounts for directly produced $\jpsi$ mesons combined with random
tracks.  Non-prompt background is due to $\jpsi$ mesons produced by
a $b$ hadron decay combined with tracks that come from either a
multibody decay of the same $b$ hadron or from hadronization.  The
mass distribution for the background is modeled by two independent
normalized negative-slope exponentials, one for the prompt and one for the
non-prompt contributions.  The PDL distribution for the prompt background is
parameterized by the resolution function described above.  The PDL distribution
for the non-prompt background is modeled by a sum of two exponential components 
for positive $ct$ and one for negative $ct$ that account for a mix of heavy 
flavor meson decays and their possible misreconstruction. The angular 
distributions for the background components are modeled by a shape similar to 
that of the signal, but with an independent set of amplitudes and phases.

\begin{figure}[htbp]
\includegraphics[scale=0.35]{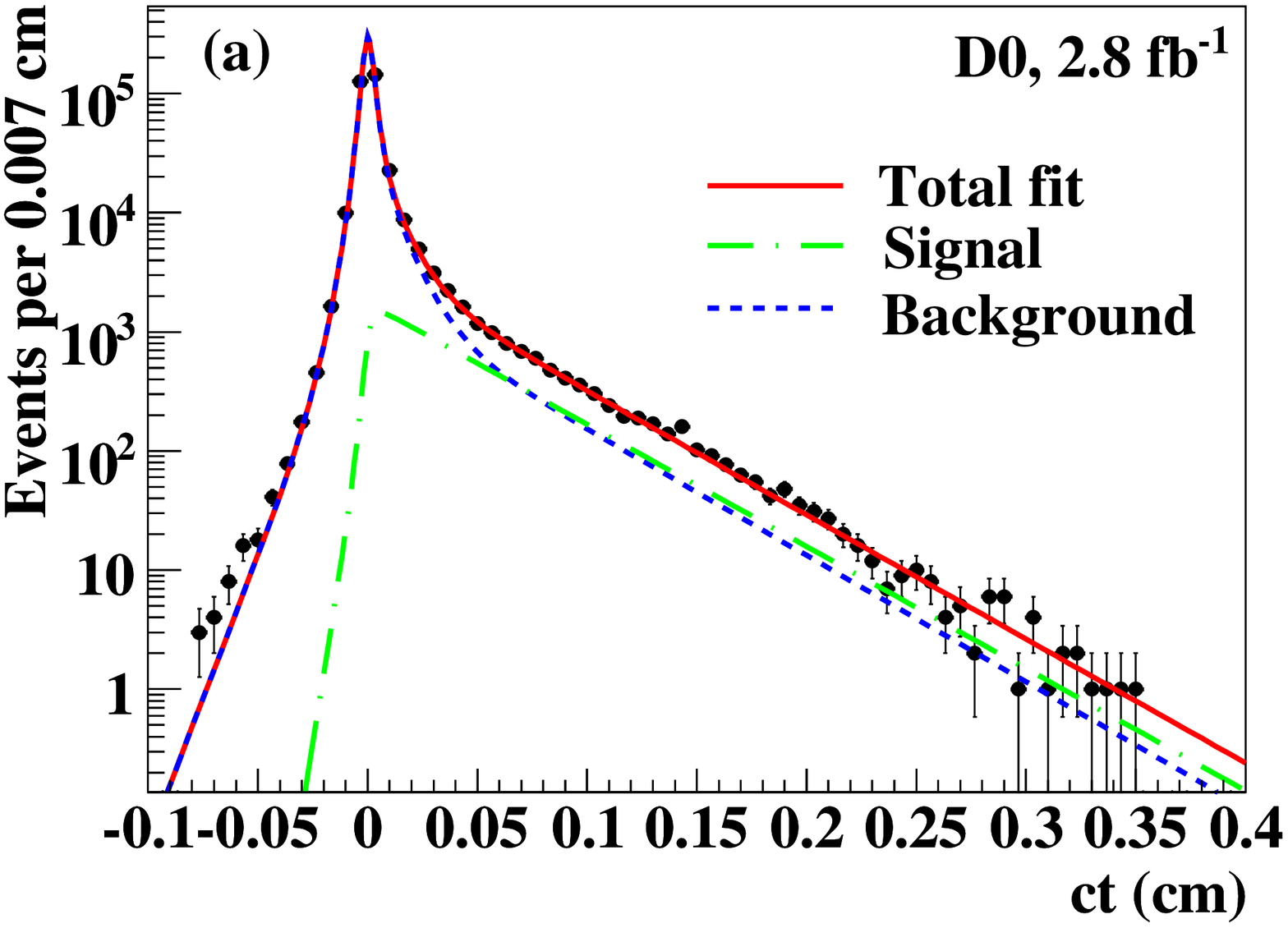}\\
\includegraphics[scale=0.35]{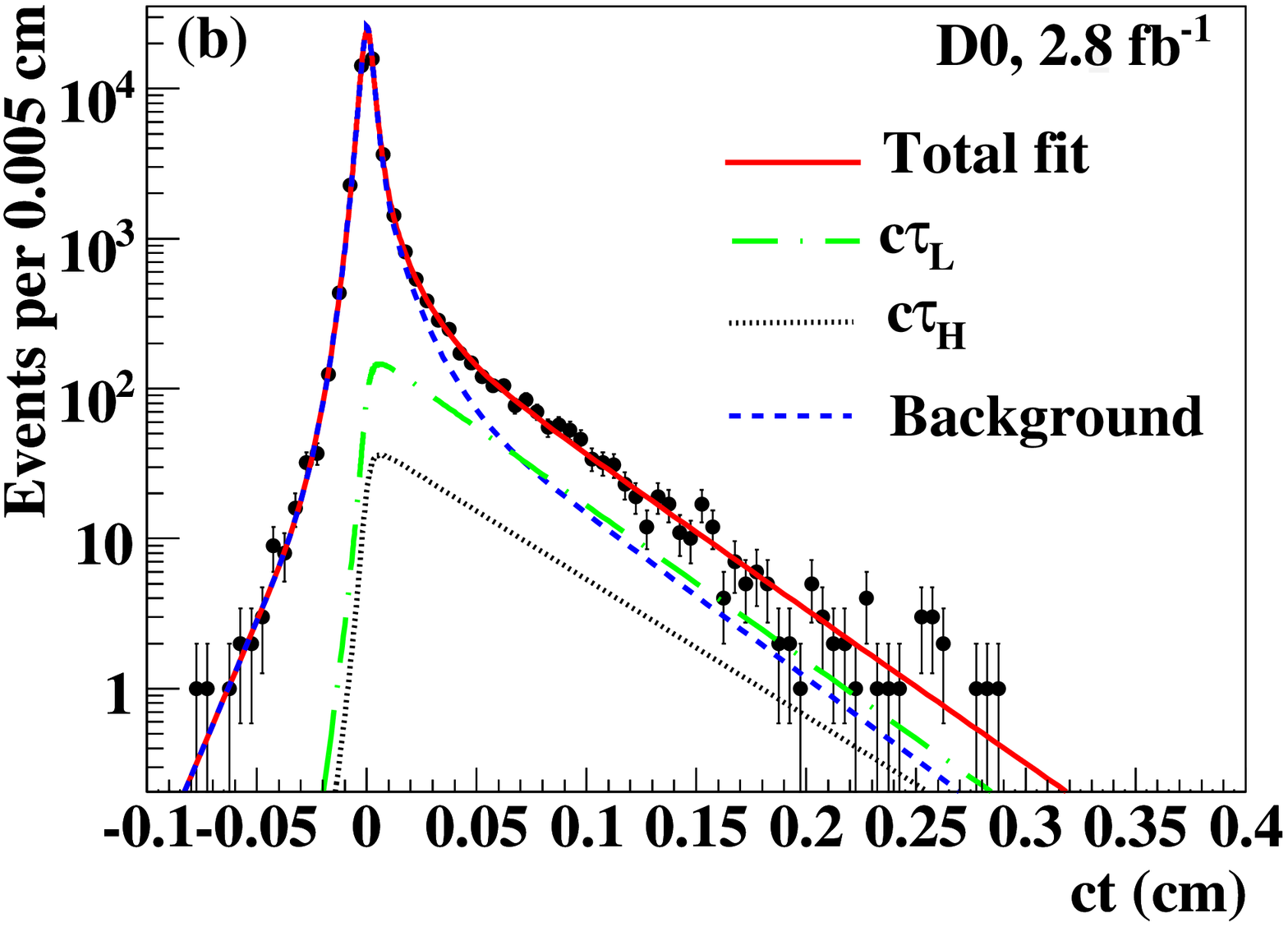}\\
\caption{ PDL distribution for selected (a) $\bd$ and (b) $\bs$ candidate events.
The points with error bars represent the data, and the curves represent the fit
projections for the total, signal, and background components.}\label{fig:pdl}
\end{figure}

The results of our measurements are summarized in Table~\ref{tab:table4}. 
Figures~\ref{fig:mass.sample.a} and~\ref{fig:pdl} show the mass and the PDL 
distributions for the $\bd$ and $\bs$ candidates, respectively, with the 
projected results of the fits.  The parameters with the strongest correlations 
are the linear amplitudes for the $\bd$, and the width difference and the mean
lifetime for the $\bs$.

\begin{table}[htbp]
\caption{Summary of measurements for the decays $\bddecay$ and $\bsdecay$. The
         uncertainties are only statistical.}\label{tab:table4}
\begin{ruledtabular}
\begin{tabular}{cccc}
 \rule{0pt}{3mm}Parameter    & $\bd$           & $\bs$                       & Units\\\hline
 \rule{0pt}{3mm}$\acero$     & $0.587\pm0.011$  & $0.555\pm0.027$           & $-$\\
 \rule{0pt}{3mm}$\all$       & $0.230\pm0.013$   & $0.244\pm0.032$ & $-$\\
 \rule{0pt}{3mm}$\duno$      & $-0.38\pm0.06$  & $-$                       & rad\\
 \rule{0pt}{3mm}$\ddos$      & $3.21\pm0.06$    & $-$                       & rad\\
 \rule{0pt}{3mm}$\dll$       & $-$               & $2.72_{-0.27}^{+1.12}$    & rad\\
 \rule{0pt}{3mm}$\tau$       & $1.414\pm0.018$   & $1.487\pm0.060$ & ps\\
 \rule{0pt}{3mm}$\dg$        & $-$              & $0.085_{-0.078}^{+0.072}$ & ps$^{-1}$\\\hline
 \rule{0pt}{3mm}$N_{sig}$& $11195\pm167$    & $1926\pm62$              & $-$\\
\end{tabular}
\end{ruledtabular}
\end{table}

\addtolength{\tabcolsep}{-0.0005\tabcolsep}
\begin{table*}
\caption{Summary of systematic uncertainties in the measurement of angular and
lifetime parameters.  The total uncertainties are given combining individual
uncertainties in quadrature.}\label{tab:tablesyst}
\begin{ruledtabular}
\begin{tabular}{lccccc|ccccc|c}
                &               &               &     $\bd$          &                    &                           &               &             &        $\bs$            &                        &                  &            \\\hline
Source   & $\acero$& $\all$  & $\duno$ (rad)&  $\ddos$ (rad)& $\tau_{d}$ (ps)& $\acero$ &$\all$ & $\dll$ (rad)& $\dg$ (ps$^{-1}$) & $\taus$ (ps)& $\tstd$\\\hline
Mass background & $-$ & $0.024$ & $0.09$ & $0.05$ & $0.030$ & $0.004$ & $0.002$ & $0.02$& $-$ &$0.021$ & $0.009$ \\
PDL resolution & $0.013$ & $0.008$ & $0.02$ & $0.03$ & $0.013$ & $0.005$ & $0.003$ & $-$ & $-$ & $0.016$ & $0.012$ \\
Fitting code & $0.001$ & $-$ & $-$ & $-$ & $0.004$ & $0.004$ & $0.014$ & $0.26$ & $0.001$ & $0.008$ & $0.003$ \\
Alignment & $-$ & $-$ & $-$ & $-$ & $0.007$ & $-$ & $-$ & $-$ & $-$ & $0.007$ & $-$ \\
\hline
Total & 0.013 & 0.025 & 0.09 & 0.06 & 0.034 & 0.006 & 0.014 & 0.26 & 0.001 & 0.028 & 0.015 \\
\end{tabular}
\end{ruledtabular}
\end{table*}
\addtolength{\tabcolsep}{-0.0005\tabcolsep}

Table~\ref{tab:tablesyst} summarizes the systematic
uncertainties in our measurements for $\bd$ and $\bs$ decays.
To study the systematic uncertainty due to the model for the mass
distributions, we vary the shapes of the mass distributions for background by
using two normalized first-order polynomials instead of the nominal two negative exponentials.
We estimate the systematic uncertainty due to the resolution on the PDL by using
one Gaussian function for the resolution model.
The fitting code is tested for the presence of biases by generating $1300$
pseudo-experiments for $\bd$ and $1000$ for $\bs$, each with the same
statistics as our data samples.  We generated the events following the 
PDL, mass, and transversity angular distributions described
above. The differences between the input and output values are quoted
as the systematic uncertainty due to the fitting.  The systematic uncertainty for 
$\dll$ reported for this source is due to an intrinsic ambiguity for this parameter 
in Eq.~(\ref{decay.distrib.bs}). The pseudo-experiments produced also cover the 
other solution for $\dll$. The contribution from the detector alignment 
uncertainty is taken from Ref.~\cite{prl.pedro}. Other potential sources of 
systematic uncertainties have been investigated and found to give negligible 
variations in the measured parameters. The systematic uncertainties for the ratio 
$\tstd$ are obtained by finding the ratio of the lifetimes for each systematic 
variation on Table~\ref{tab:tablesyst} and taking the difference between this 
value and the nominal ratio.

In conclusion, we have measured the angular and lifetime parameters for the
time-dependent angular untagged decays $\bddecay$ and $\bsdecay$,
the lifetime ratio of both $B$ mesons, and the width difference $\dg$ for the 
$\bs$ meson. From the measured lifetime parameters $\taus$ and $\tau_{d}$ we 
obtain the ratio 
$\tstd=1.052\pm0.061\mathrm{\thinspace(stat)}\pm0.015\mathrm{\thinspace(syst)}$  
which is consistent with the theoretical prediction~\cite{theorylifetime} and 
previous measurements~\cite{pdg1}. The measurement of the width difference
$\dg=0.085_{-0.078}^{+0.072}\mathrm{\thinspace(stat)}\pm0.006\mathrm{\thinspace(syst)}$~\thinspace{ps}$^{-1}$
is consistent with the theoretical prediction \cite{theorylifetime,lenz} and with 
the value reported in Refs. \cite{pdg1,cdf}. D0 also has a measurement of $\dg$ 
in a flavor-tagged analysis of $\bsdecay$ in Ref.~\cite{tagged.d0}.

Our measurements for the linear polarization amplitudes for the $\bd$, taking 
into account the interference between the $K\pi$ $S$-wave and $P$-wave, are
$\acero=0.587\pm0.011\mathrm{\thinspace(stat)}\pm0.013\mathrm{\thinspace(syst)}$ and
$\all=0.230\pm0.013\mathrm{\thinspace(stat)}\pm0.025\mathrm{\thinspace(syst)}$; and for $\bs$:
$\acero=0.555\pm0.027\mathrm{\thinspace(stat)}\pm0.006\mathrm{\thinspace(syst)}$, and
$\all=0.244\pm0.032\mathrm{\thinspace(stat)}\pm0.014\mathrm{\thinspace(syst)}$ are 
consistent and competitive with those reported in the literature \cite{babar,belle,pdg1}. 
Our measurement of the strong phases $\duno$ and $\ddos$ indicates the presence of 
final-state interactions for the decay $\bddecay$~\cite{Fleischer} since 
$\duno=-0.38\pm0.06\mathrm{\thinspace(stat)}\pm0.09\mathrm{\thinspace(syst)}$\thinspace{rad} 
is $3.5\sigma$ away from zero, where $\sigma$ is the total uncertainty. From the 
comparison of the measured amplitudes and strong phases~\cite{dll} for both decays 
we conclude that they are consistent with being equal for $\bd$ and $\bs$ and hence 
there is no evidence for a deviation from flavor SU(3) symmetry. In our sample we 
find that the $K\pi$ $S$-wave intensity, as described in Ref.~\cite{babar}, is 
$(4.0 \pm 1.0)\%$.

%
We thank the staffs at Fermilab and collaborating institutions, 
and acknowledge support from the 
DOE and NSF (USA);
CEA and CNRS/IN2P3 (France);
FASI, Rosatom and RFBR (Russia);
CNPq, FAPERJ, FAPESP and FUNDUNESP (Brazil);
DAE and DST (India);
Colciencias (Colombia);
CONACyT (Mexico);
KRF and KOSEF (Korea);
CONICET and UBACyT (Argentina);
FOM (The Netherlands);
STFC (United Kingdom);
MSMT and GACR (Czech Republic);
CRC Program, CFI, NSERC and WestGrid Project (Canada);
BMBF and DFG (Germany);
SFI (Ireland);
The Swedish Research Council (Sweden);
CAS and CNSF (China);
and the
Alexander von Humboldt Foundation (Germany).


\begin{thebibliography}{99}
%
\bibitem[a]{alton}
Visitor from Augustana College, Sioux Falls, SD, USA.
\bibitem[b]{burdin}
Visitor from The University of Liverpool, Liverpool, UK.
\bibitem[c]{gershtein}
Visitor from Rutgers University, Piscataway, NJ, USA.
\bibitem[d]{hensel,meyer,park,quadt}
Visitor from II. Physikalisches Institut, Georg-August-University,
  G{\"o}ttingen, Germany.
\bibitem[e]{luna-garcia}
Visitor from Centro de Investigacion en Computacion - IPN,
  Mexico City, Mexico.
\bibitem[f]{podesta-lerma}
Visitor from ECFM, Universidad Autonoma de Sinaloa, Culiac\'an, Mexico.
\bibitem[g]{voutilainen}
Visitor from Helsinki Institute of Physics, Helsinki, Finland.
\bibitem[h]{weber}
Visitor from Universit{\"a}t Bern, Bern, Switzerland.
\bibitem[i]{wenger}
Visitor from Universit{\"a}t Z{\"u}rich, Z{\"u}rich, Switzerland.
\bibitem[\ddag]{deceased}
Deceased.

%
\vskip 0.25cm

  \bibitem{browder}T.E.~Browder, K.~Honscheid, and D.~Pedrini, Annu. Rev. Nucl. Part. Sci. {\bf 46}, 395 (1996).
  \bibitem{Fleischer}A.S.~Dighe, I.~Dunietz, and R.~Fleischer, Eur. Phys. J. {\bf C6}, 647 (1999), and references therein.
  \bibitem{charge.conjugation}Unless explicitly stated, the appearance of a specific charge state will also imply
         its charge conjugate throughout the paper.
  \bibitem{franco} E. Franco {\sl et al.}, Nucl. Phys. {\bf B633}, 212 (2002).
  \bibitem{theorylifetime}A. Lenz, arXiv:0802.0977 [hep-ph] (2008).
  \bibitem{pdg1}W.-M. Yao {\sl et al.} (Particle Data Group), J. Phys. {\bf G 33}, 1 (2006) and 2007 partial
  update for the 2008 edition.
  \bibitem{run2det} V. Abazov {\sl et al.} [D0 Collaboration], Nucl. Instr. and Meth. Phys. Res. A {\bf 565}, 463 (2006).
  \bibitem{tagged.d0}V. Abazov {\sl et al.} [D0 Collaboration], arXiv:0802.2255 [hep-ex] (2008), submitted to Phys.~Rev.~Lett.
  \bibitem{prl.daria} V. Abazov {\sl et al.} [D0 Collaboration], Phys. Rev. Let. \textbf{98}, 121801 (2007).
  \bibitem{prl.eduard}V. Abazov {\sl et al.} [D0 Collaboration], Phys. Rev. Lett.  \textbf{94}, 102001 (2005).
  \bibitem{prl.pedro} V. Abazov {\sl et al.} [D0 Collaboration], Phys. Rev. Lett. \textbf{94}, 042001 (2005).
  \bibitem{amplitudes} Throughout the paper, if not explicit dependence on time is stated, we denote $A_{i}(0)\equiv A_{i}$ 
           for $i=\{0,\parallel,\perp\}$.
  \bibitem{lenz}A. Lenz and U. Nierste, JHEP {\bf 06}, 072 (2007).
  \bibitem{babar} B. Aubert {\sl et al.} [BaBar Collaboration], Phys. Rev. D \textbf{71}, 032005 (2005).
  \bibitem{belle} K. Abe {\sl et al.} [Belle Collaboration], Phys. Rev. Lett. \textbf{95}, 091601 (2005).
  \bibitem{cdf} T.~Aaltonen {\sl et al.}, [CDF Collaboration], Phys. Rev. Lett. {\bf 100}, 121803 (2008).
  \bibitem{dll} Using the relation between these phases we obtain
             $\delta_{\parallel,\bd}~=~3.59\pm0.06\pm0.09$\thinspace\thinspace{rad}.
\end{thebibliography}
\end{document}